\documentclass[12pt,preprint2]{aastex}
\usepackage{emulateapj5}
\usepackage{onecolfloat5}

\usepackage{natbib}

\newcommand{\rxj}{RX~J0720.4$-$3125}
\newcommand{\rxjw}{RX~J1856.5$-$3754}
\newcommand{\chandra}{\textit{Chandra}}

\newcommand{\hst}{\textit{HST}}
\newcommand{\rosat}{\textit{ROSAT}}
\newcommand{\xmm}{\textit{XMM}}
\newcommand{\psr}{PSR~J1814$-$1744}
\newcommand{\opsr}{PSR~J2144$-$3944}

\newcommand{\beq}{\begin{equation}}
\newcommand{\eeq}{\end{equation}}
\newcommand{\mc}{\multicolumn}
\newcommand{\us}{\ensuremath{\mbox{ }\mu\mbox{s}}}      
\newcommand{\expnt}[2]{\ensuremath{#1 \times 10^{#2}}}   
\newcommand{\gsim}{\gtrsim}
\newcommand{\lsim}{\lesssim}

\shorttitle{The Spectrum of \rxj}
\shortauthors{Kaplan et al.}

\slugcomment{Accepted by ApJ}

\begin{document}
\twocolumn[
\title{The Nearby Neutron Star \rxj\ from Radio to X-rays}
\author{D.~L.~Kaplan\altaffilmark{1},
M.~H.~van~Kerkwijk\altaffilmark{2}, H.~L.~Marshall\altaffilmark{3},
B.~A.~Jacoby\altaffilmark{1}, S.~R.~Kulkarni\altaffilmark{1} \and
  D.~A.~Frail\altaffilmark{4}}

\begin{abstract}
We present radio, optical, ultraviolet, and X-ray observations of the
isolated, thermally-emitting neutron star \rxj\ using the Parkes radio
telescope, the Very Large Array, the \textit{Hubble Space Telescope},
and the \textit{Chandra X-ray Observatory}.  From these data we show
that the optical/UV spectrum of \rxj\ is not well fit by a
Rayleigh-Jeans tail as previously thought, but is instead best fit by
either a single non-thermal power-law or a combination of a
Rayleigh-Jeans tail and a non-thermal power-law.  Taken together with
the X-ray spectrum, we find the best model for \rxj\ to be two
blackbodies plus a power-law, with the cool blackbody implying a
radius of 11--13~km at an assumed distance of 300~pc.  This is similar
to many middle aged ($10^{5-6}$~yr) radio pulsars such as
PSR~B0656+14, evidence supporting the hypothesis that \rxj\ is likely
to be an off-beam radio pulsar.  The radio data limit the flux at
1.4~GHz to be $<0.24$~mJy, or a luminosity limit of $4\pi d^{2}F <
\expnt{3}{25} d_{300}^{2}\mbox{ ergs s}^{-1}$, and we see no sign of
extended nebulosity, consistent with expectations for a pulsar like
\rxj.
\end{abstract}


]
\altaffiltext{1}{Department of Astronomy, 105-24
California Institute of Technology, Pasadena, CA 91125, USA; dlk@astro.caltech.edu}
\altaffiltext{2}{Dept. of Astronomy \& Astrophysics, 60 St George St.,
  Toronto, ON, M5S 3H8, Canada}
\altaffiltext{3}{Center for Space Research, Massachusetts Institute of
Technology, Cambridge, MA 02139, USA}
\altaffiltext{4}{National Radio Astronomy Observatory,
  P.O. Box O, Socorro, NM 87801, USA}

\section{Introduction}
Thermally-emitting neutron stars have been the targets of many
observations recently, as these sources can potentially reveal the
equation of state (EOS) of neutron stars, and thereby explore nuclear
physics in realms inaccessible from laboratories \citep{lp00}.  To
obtain the EOS from the spectrum seems simple: determine the effective
angular size from spectral fits, multiply by the distance (obtained
from other means), and one has the apparent radius.  This radius can
be converted into the physical radius through use of mass.  The radius
is the crucial quantity in differentiating between EOS, as most EOS
predict a distinctive but small range of radii for a large range of
masses.  However, in order to use a neutron star to determine the EOS,
one needs to (1) be certain that the radiation is from the surface,
and (2) have a thorough understanding of this emission.  Radio pulsars
and accreting binaries are unsuitable since non-thermal magnetospheric
emission or emission from the accreting material far exceeds the
surface emission.

Therefore the identification of the nearest neutron stars by \rosat\
was a major advance in the field (see reviews by \citealt{motch01} and
\citealt{ttzc00}).  Most of these sources do not have significant
non-thermal emission, so they are prime targets for studies leading to
the EOS.  The closest of these sources, \rxjw, has been the subject of
much inquiry lately for just this purpose \citep[e.g.,][]{wl02,dmd+02,br02}.

\rxj\ was discovered by \citet{hmb+97} as a soft ($kT \sim 80$~eV),
bright X-ray source in the \rosat\ All-Sky Survey --- the second
brightest neutron star that is not a radio pulsar.  Given its very low
hydrogen column density ($N_H\sim \expnt{1}{20}\mbox{ cm}^{-2}$),
nearly sinusoidal 8.39-s pulsations, relatively constant X-ray flux,
and very faint ($B=26.6$~mag), blue optical counterpart
\citep{kvk98,mh98}, it was classified as a nearby, isolated,
thermally-emitting neutron star.  It is perhaps the second closest
source (next to \rxjw) that does not show significant non-thermal
emission.  While originally thought to be an accreting source or
possible an old magnetar, recent timing measurements limit the
original magnetic field to be smaller than $10^{14}$~G, eliminating
the magnetar hypothesis.  However, the discoveries of radio pulsars
with periods longer than 4~s (\citealt{ckl+00}; \citealt*{ymj99}) have
led to the suggestion that \rxj\ is instead an off-beam radio pulsar,
likely with a magnetic field at the high end of the radio-pulsar range
\citep[$B\sim 10^{13}$~G; ][]{kkvkm02,zhc+02}.

The spectrum emerging from a thermally-emitting neutron star depends
significantly on the composition of the surface
\citep[e.g.,][]{romani87}.  In the past, three models have generally
been considered.  The first, a blackbody, is simple but not physically
motivated.  Next, light element (H or He, possibly due to accretion
from the ISM) atmosphere have few features, all in the (extreme) ultraviolet,
and peak, for a given temperature, at a substantially higher energy
than a black-body.  Finally, heavy element (Fe, Ni) atmospheres peak
at a similar location to blackbodies but have many spectral features
at a variety of wavelengths.

None of these models can reproduce the X-ray and optical data for
\rxj\ and \rxjw\ \citep{pwl+02,wl02,dmd+02,pmm+01}.  The X-ray data
are well fit by blackbodies, but these blackbodies underpredict the
optical flux.  The H/He models also match the general shape, but they
overpredict the optical flux and the implied radii are larger than is
possible for a neutron star.  The Fe models have too many lines and
edges to match the X-ray spectra.  Consequently, the current
generation of single-component models have been unsuccessful in
fitting both the X-ray and optical fluxes while having radii
consistent with those of neutron stars (\citealt{pwl+02};
\citealt*{kvka02}).  We see, though, how both X-ray and optical data
are required to fully constrain the models.

Motivated thus, we present observations of \rxj\ from radio to X-ray
wavelengths aimed at determining its spectral energy distribution and
from that its underlying properties (composition, magnetic field, and
geometry).  In Section~\ref{sec:opt} we present new optical/UV data
from the \textit{Hubble Space Telescope}, and undertake detailed
modeling of the optical/UV spectrum.  In Section~\ref{sec:xray} we
present spectroscopic data from the \textit{Chandra X-ray
Observatory}, and in Section~\ref{sec:radio} we present searches for
radio sources (both persistent and pulsating) with the Very Large
Array and the Parkes radio telescope.  Finally, in
Section~\ref{sec:discuss} we discuss the spectrum of \rxj, and present
our conclusions in Section~\ref{sec:conc}.  In the following, all
radii refer to the radiation radius as observed at infinity,
$R_{\infty}$ ($R_\infty=R_{\rm phys}/\sqrt{1-2GM/R_{\rm phys}c^2}$,
where $R_{\rm phys}$ is the physical radius).

\section{Optical and UV Data}
\label{sec:opt}
\subsection{Observations and Analysis}
\label{sec:obs}
We observed \rxj\ with the Space Telescope Imaging Spectrometer (STIS)
aboard the \textit{Hubble Space Telescope} (\hst) four times, covering
wavelengths from 125~nm to 900~nm; the observations are summarized
in Table~\ref{tab:obs}.

\begin{deluxetable}{l c c c c c c c}
\tabletypesize{\small}
\tablecaption{Summary of STIS Observations\label{tab:obs}}
\tablewidth{0pt}
\tablehead{
\colhead{Detector/Filter} & \colhead{Date} & \colhead{Exposure}
&\colhead{$\langle\lambda\rangle$\tablenotemark{a}} &
\colhead{$\langle A_{\lambda}/A_{V}\rangle$\tablenotemark{a}} &
\colhead{Ap.\ Corr.\tablenotemark{b}} & \colhead{$\Delta Z_{\rm mag}$\tablenotemark{c}}
& \colhead{Magnitude\tablenotemark{d}}  \\ 
& \colhead{(UT)} & \colhead{(sec)} & \colhead{(\AA)} & \colhead{(mag)} &
\colhead{(mag)} & \colhead{(mag)} &\colhead{(mag)} \\
}
\startdata
 & & & \mc{5}{c}{Calculated for $\alpha_{\nu}=2.00$ and
  $A_{0}=0.13$~mag.\tablenotemark{e}} \\
CCD/50CCD & 2001-Jul-16 & 5342 & 5148 & 1.56& 0.102 &\nodata &$26.68\pm0.10$ \\
NUV MAMA/25Qtz & 2001-Aug-05 & 5500 & 2286& 2.58 &0.276& 0.029 &$23.82\pm0.14$ \\
FUV MAMA/25SrF$_{2}$ & 2002-Jan-28 & 4800 & 1447&  2.80&0.340 &0.087 &
$21.97\pm0.11$ \\
FUV MAMA/25MAMA & 2002-Feb-13 & 3850 & 1360&  2.97&0.353 &0.078 &$21.6\pm0.2$ \\[0.1in]
 & & & \mc{5}{c}{Calculated for $\alpha_{\nu}=1.40$ and
  $A_{0}=0.10$~mag.\tablenotemark{f}} \\
CCD/50CCD & \nodata & \nodata & 5367 & 1.41& 0.102 &\nodata &$26.68\pm0.10$ \\
NUV MAMA/25Qtz &\nodata & \nodata  & 2310& 2.56 &0.270& 0.029 &$23.83\pm0.14$ \\
FUV MAMA/25SrF$_{2}$ &\nodata & \nodata  & 1450&  2.79&0.339 &0.088 &
$21.96\pm0.11$ \\
FUV MAMA/25MAMA &\nodata & \nodata  & 1365 &  2.95&0.352 &0.079 &$21.7\pm0.2$ \\

\enddata \tablenotetext{a}{Effective wavelength and normalized
extinction; see Appendix~A of \citet{vkk01}.}
\tablenotetext{b}{Aperture correction from a radius of $0\farcs5$ to
infinity.  See \S~\ref{sec:opt} for details.}
\tablenotetext{c}{Change in the zero-point magnitude due to
degradation of the MAMA detectors.}  \tablenotetext{d}{Magnitude in
the STMAG system, with $m_{\rm ST}=-21.1-2.5\log_{10} F_{\lambda}$,
corrected to infinite aperture.}  \tablenotetext{e}{Appropriate for
the RJ fit in Table~\ref{tab:fit}.}  \tablenotetext{f}{Appropriate for
all fits except the RJ fit in Table~\ref{tab:fit}.}
\tablecomments{All filter curves were taken from the \texttt{synphot}
database.  $\lambda_{0}=4500$~\AA.}
\end{deluxetable}

\subsubsection{Optical Data}

\begin{figure}[t]
\plotone{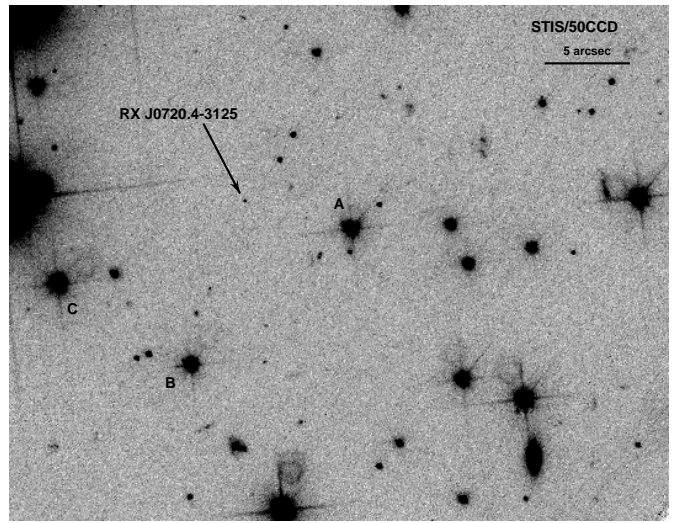}
\caption{STIS optical image of \rxj.  North is up, East to the left.
  The scale bar in the upper right indicates $5\arcsec$.  \rxj\ is
  indicated by the arrow, and sources A, B, and C from \citet{hmb+97}
  are also marked.}
\label{fig:stis}
\end{figure}

The optical data (50CCD mode) consist of eight unfiltered CCD observations
taken in a four-point dither pattern.  We assembled the
images using the drizzle algorithm \citep{fh02} giving a plate-scale
of $25.3\mbox{ mas pixel}^{-1}$.  We show the stacked image in
Figure~\ref{fig:stis}.

We performed standard aperture photometry on the stacked STIS image
using \texttt{IRAF}'s \texttt{daophot} package.  The sky level was
estimated using an annulus from $0\farcs75$--$1\farcs00$.  The source
flux was measured within an aperture of radius $0\farcs5$.  While
there were no aperture corrections strictly appropriate for a source
with the color of \rxj\ ($B-V \approx -0.3$~mag), we used the bluest
of the color-dependent aperture corrections available (T.~Brown 2002,
personal communication) to correct the flux to an infinite aperture.
We estimate that the aperture correction introduces an uncertainty of
$<0.02$~mag, as blue sources like \rxj\ have less scattered light than
redder sources and therefore the aperture corrections are better
determined \citep[see][]{kkvk02}.  Given the very wide 50CCD bandpass 
 (${\rm FWHM}\approx 441$~nm) a single zero-point flux is
not appropriate for all source spectra. Therefore, as a first order
estimate, we calculated the
zero-point flux at the mean wavelength of the filter (given in
Table~\ref{tab:obs}), assuming an input spectrum with
$F_{\lambda}\propto \lambda^{-4}$ and $A_{V} \approx 0.1$~mag,
appropriate for \rxj\ \citep{kvk98}.

For astrometric purposes we used 10-s and 60-min $R$-band images taken
with the Low-Resolution Imaging Spectrograph \citep{o+95} on the 10-m
Keck~I telescope (the observations are described in \citealt{kvk98}).
We determined the centroids of 231 stars from the Guide Star Catalog
version 2.2 (GSC-2.2) on the 10-s image, rejecting 50 objects that
were overexposed or appeared to be incorrectly identified.  The pixel
coordinates were corrected for instrumental distortion using a cubic
radial distortion
function\footnote{See \url{http://alamoana.keck.hawaii.edu/inst/lris/coordinates.html}.},
and we then fit for the zero-point position, rotation, and plate-scale
of the image.  The rms was $0\farcs12$ in each coordinate.  We then
fit the 70 non-saturated stars from the 10-s image to the 60-min
composite image, again correcting for instrumental distortion.  This
fit had rms of $0\farcs024$ in each coordinate.  Finally, we performed
a fit using 25 stars from the 60-min image to determine the
zero-point, scale, and rotation of the STIS image (which had been
corrected for distortion during the drizzle process), giving an rms of
$0\farcs025$ in each coordinate.  Overall, the STIS image is tied to
the frame of the GSC-2.2 with uncertainty of $0\farcs01$ in each
coordinate, or tied to the International Coordinate Reference Frame
(ICRF) with uncertainty of about $0\farcs2$ in each coordinate.  The
final position for \rxj\ is (J2000, epoch MJD~52106) $\alpha=07^{\rm
h}20^{\rm m}24\fs961$, $\delta=-31\degr25\arcmin50\farcs21$.  This
supersedes the position of \citet{kvk98}, as the significance of the
new detection is far higher and the position is directly tied to the
ICRF. The \hst\ position is consistent with the X-ray position which
has uncertainties of $\approx 0\farcs6$ in each coordinate \citep{kkvkm02}.

\subsubsection{UV Data}
For the UV MAMA data, we corrected the arrival times of the data to
the Solar System barycenter using the \texttt{stsdas} task
\texttt{odelaytime}.  We then filtered the photon lists for the FUV
data for periods of high background.  These occurred at the beginning
of each orbit as \hst\ was going into the Earth's shadow.  Using a
background annulus from $3\farcs0$--$3\farcs5$, we estimated the
median background level from a light-curve binned to 50-s and only
used the data where the background level was within $\pm 3\,\sigma$ of
the median.  We found that the background was noticeably high only for
the FUV MAMA/25SrF$_{2}$ data, where we have retained 4800~s of the
original 5500~s.

\begin{deluxetable}{c c c}
\tablecaption{Source and Background Counts for STIS UV Data\label{tab:uv}}
\tablewidth{0pt}
\tablehead{
\colhead{Detector/Filter} & \mc{2}{c}{Counts} \\\cline{2-3}
 & \colhead{Source} & \colhead{Background} \\
}
\startdata
NUV MAMA/25Qtz & $3534\pm 59$ & $2788\pm15$\\
FUV MAMA/25SrF$_{2}$ & $611 \pm 25$ & $128\pm 3$\\
FUV MAMA/25MAMA & $14502\pm 120$ & $13788\pm58$\\
\enddata
\tablecomments{No aperture corrections have been applied.  Background
  counts are normalized to the same area as the source counts.}
\end{deluxetable}

We then performed aperture photometry on the raw photon data.  The
source flux was computed using a radius of $0\farcs5$, where the
signal-to-noise was relatively high and the aperture corrections were
well-defined.  The raw source and background counts are given in
Table~\ref{tab:uv}.  The STIS MAMAs have wide point-spread-functions
(psfs), with substantial flux beyond $0\farcs5$, and therefore
aperture corrections are particularly important.  We took the
monochromatic aperture corrections for the MAMAs (T.~Brown 2002,
personal communication) convolved with the expected source spectrum
and the filter throughputs to compute aperture corrections for each
filter, which we give in Table~\ref{tab:obs}. Another issue with the
STIS MAMAs is that the sensitivity changes with time \citep{swb+02} at
the level of a few percent per year.  Using a wavelength-dependent fit
to the sensitivity changes (D.~Stys 2002, personal communication;
R.~Diaz-Miller 2002, personal communication; these corrections are now
incorporated into the STIS pipeline), we computed weighted zero-point
corrections for the NUV and FUV MAMA data that are listed in
Table~\ref{tab:obs}.  These corrections were small, less than 0.1~mag.
As with the optical data, the zero-point fluxes were calculated at the
mean wavelengths of the filters.

The STIS MAMA data are time-tagged with $125\us$ resolution, but we
could not detect the 8.39-s periodicity present in the X-rays.  We
find 90\% confidence upper limits on the Fourier power (normalized to
have unity mean and rms) present at the X-ray period (given by
\citealt{kkvkm02}) of 5.3 and 5.9, for the NUV and FUV/SrF$_{2}$ data
sets, respectively.  These translate into limits on the rms pulsed
fraction of 0.16 and 0.19, respectively.

\subsection{Spectral Fits}
\subsubsection{Power-law Fits}
\label{sec:pl}
We already know from \citet{kvk98} that the optical spectrum of \rxj\
is roughly approximated by the Rayleigh-Jeans portion of a blackbody
curve: $F_{\lambda}\propto \lambda^{-4}$.  We can now investigate this more
quantitatively and over a wider range of wavelengths.  Here we use the
optical and UV data presented in this paper as well as the $B$ and $R$
photometry from \citet{kvk98}; for the data from \citet{kvk98}, we
used the zero-point fluxes from \citet*{bcp98}.  

Our spectral fitting followed \citet{vkk01}.  We fit the data with a
sum of extincted power-law (PL) of the form 
\beq 
F_{\lambda} =
F_{0}\left(\frac{\lambda}{\lambda_{0}}\right)^{-(2+\alpha_{\nu})}
10^{-A_{0}A^{\prime}_{\lambda}/2.5}, 
\eeq 
where $\alpha_{\nu}$ is the
spectral index\footnote{From this definition we also have the more
standard $F_{\nu} \propto \nu^{\alpha_{\nu}}$.}, $F_{0}$ is the
observed flux at the reference wavelength $\lambda_{0}$, $A_{0}$ is
the extinction at $\lambda_{0}$, and $A^{\prime}_{\lambda}$ is the
normalized extinction at wavelength $\lambda$ ($A^{\prime}_{\lambda}
\equiv A_{\lambda}/A_{0}$). We use the reddening curve of
\citet*{ccm89}, with corrections to the optical and UV portions from
\citet{o94}.   We chose $\lambda_{0}=4500\mbox{ \AA}$ ($A_{0} = 1.29A_{V}$,
or $A_{0} \approx A_{B}$), as this is the mean wavelength of the data.

For the fit, we calculated likelihood values as we varied $F_{0}$,
$A_{0}$, and $\alpha_{\nu}$.  However, as the spectral shape changes (i.e.\
variations in $A_{0}$ and $\alpha_{\nu}$) the aperture corrections,
reference wavelengths, extinctions, and zero-point corrections (from
Table~\ref{tab:obs}) also change (these changes are most
significant for the STIS/50CCD data).  Therefore, during the
iterations of the fitting, we recomputed all of the spectrum-dependent
quantities for each combination of $A_{0}$ and $\alpha_{\nu}$.

The results of the fitting for the two basic models ---
Rayleigh-Jeans\footnote{For blackbodies of the temperatures considered
here, the Rayleigh-Jeans approximation holds at the shortest UV
wavelength used to better than 4\% --- considerably smaller than the
measurement error.}  (RJ) and unconstrained power-law (PL) are given
in Table~\ref{tab:fitnoA}.  These fits give values of the extinction
$A_{0} \sim 0.5$~mag.  This is much higher than expected from other
observations.  From the X-ray spectrum of \rxj, we know that the
hydrogen column density is $N_{H} \approx \expnt{1.3}{20}\mbox{
cm}^{-2}$, which implies $A_{V} \approx 0.07$~mag \citep{ps95} or
$A_{0} \approx 0.09$~mag.  \citet{kvk98} have placed an upper limit on
the reddening of $E_{B-V} < 0.04$, as \rxj\ is in the foreground of
the open cluster Collinder~140, which implies $A_{0} < 0.10$ (for the
standard ratio of $R_{V}=3.2$).  We therefore expect small values of
the extinction: $A_{0} \lsim 0.15$~mag, allowing for uncertainties in
$N_{H}$, $R_{V}$, and the relation between $N_{H}$ and $A_{V}$.  From
this we can reject the models in Table~\ref{tab:fitnoA}.

\begin{deluxetable}{c c c }
\tabletypesize{\small}
\tablecaption{Fits to \rxj\ Optical/UV Data\label{tab:fitnoA} with
 $A_{0}$  Unconstrained}
\tablewidth{0pt}
\tablehead{
\colhead{Parameter} & \mc{2}{c}{Type of Fit} \\  \cline{2-3}
 & \colhead{PL} &  \colhead{RJ}\\
}
\startdata
$A_{0}$ (mag) & 0.47(15) & 0.59(6) \\
\tableline
$\alpha_{\nu}$\tablenotemark{a} & 1.9(2) &  2 \\
$F_{0}$ $( \times 10^{-19} \mbox{ ergs s}^{-1}\mbox{ cm}^{-2}\mbox{ \AA}^{-1})$ 
 & 2.5(6) & 3.0(3) \\
$B_{0}$ (mag)\tablenotemark{b}  & 26.0(2) & 25.82(10)\\
\tableline
$\chi^{2}$ & 1.0 & 1.1  \\
DOF & 3& 4 \\
$\chi^{2}/{\rm DOF}$ & 0.3& 0.3 \\
\enddata
\tablenotetext{a}{The spectral index such that $F_{\lambda} \propto \lambda^{-(2+\alpha_{\nu})}$.}
\tablenotetext{b}{$B$-band Vega-magnitude corresponding to $F_{0}$.}
\tablecomments{Numbers in parentheses are 68\% confidence limits in
  the last digit(s).
  Values without confidence limits were held fixed for the fit.
  Values with subscript $0$ are for $\lambda_{0}=4500$~\AA.}
\end{deluxetable}

Therefore we constrained $A_{0}$ from the information above.  To
formally include this in 
our fit, we performed a maximum likelihood fit with the following
prior distribution for $A_0$: \beq f_{A_{0}}(A_{0}) = {\cal
N}_{6}(A_{0} | 0.09\mbox{ mag}, 0.06\mbox{ mag}),
\label{eqn:prior}
\eeq
where ${\cal N}_{n}(x|\mu, \sigma)$ is a generalized Gaussian
distribution of degree $n$ ($n$ is even) with the form:
\begin{eqnarray}
{\cal N}_{n}(x | \mu,\sigma) &\equiv &\frac{n}{2\sqrt{2} \pi \sigma}
\Gamma\left(\frac{n-1}{n} \right) \sin \left(\frac{\pi}{n}\right)
\times \nonumber \\
 & & 
\exp \left[ - \left( \frac{(x-\mu)}{\sqrt{2} \sigma}\right)^{n} \right].
\label{eqn:gengauss}
\end{eqnarray}
For $n=6$, this distribution essentially requires that $A_{0}$ be between 0~mag
and 0.18~mag.  We could have used a uniform prior with
$A_{0}=0$--$0.15$~mag, but the sharp edges of this distribution can make
for discontinuities in the resulting posterior distributions, so we opted
for Equation~\ref{eqn:prior}, which is a  smoothed version of the
uniform distribution.

We consider four power-law models for the fit.  The first, given in
the PL column in Table~\ref{tab:fit}, is a single power-law fit to the
optical/UV data alone.  The second, given in the RJ column in
Table~\ref{tab:fit}, is a single power-law fit to the optical/UV data,
but where the power-law index is that of a Rayleigh-Jeans tail
($\alpha_{\nu}=2$).  The third is a fit where there are two
power-laws, given in the PL+RJ column in Table~\ref{tab:fit}: one has
an unconstrained index, and the other has $\alpha_{\nu}=2$.  Finally,
the fourth fit, given in the PL+X-ray column in Table~\ref{tab:fit},
has a Rayleigh-Jeans power-law present, but its normalization is set
by the X-ray fit (\S~\ref{sec:xray}; the uncertainties in the X-ray
flux extrapolated to optical/UV wavelengths is $\approx 10$\%).

\begin{figure}[b]
\plotone{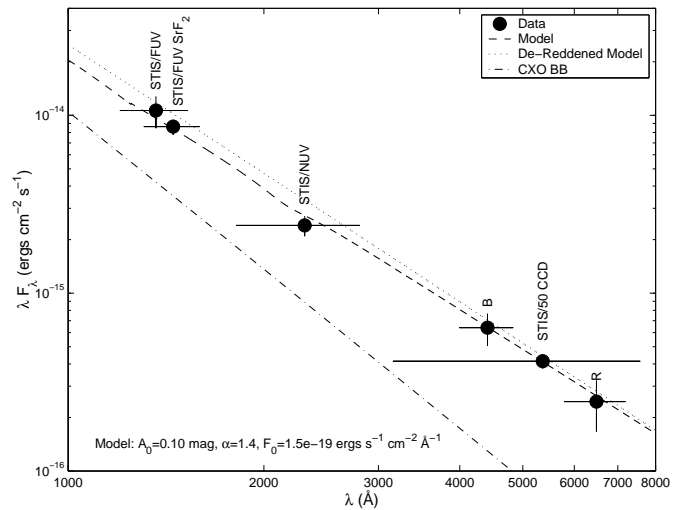}
\caption{Optical/UV spectrum of \rxj, with data from \citet{kvk98} and
this paper.  The PL fit is shown, with: the best-fit model (dashed
line), the best-fit model corrected for extinction (dotted line) and
the extrapolation of the \chandra\ blackbody fit (dash-dotted line;
``CXO BB''; see \S~\ref{sec:xray}).  The different bands are labeled.}
\label{fig:spec}
\end{figure}

\begin{deluxetable}{c c c c c}
\tabletypesize{\small}
\tablecaption{Fits to \rxj\ Optical/UV Data\label{tab:fit} with
  $A_{0}$ Constrained}
\tablewidth{0pt}
\tablehead{
\colhead{Parameter} & \mc{4}{c}{Type of Fit} \\  \cline{2-5}
 & \colhead{PL} &  \colhead{RJ}  & \colhead{PL+RJ} &
\colhead{PL+X-ray}\\
}
\startdata
$A_{0}$ (mag) & 0.10(6) & 0.19(2)&  0.10(4) & 0.09(4)\\
\tableline
$\alpha_{\nu}$\tablenotemark{a} & 1.40(4) & 2& 0.3(7) & 1.14(12)\\
$F_{0}$ $( \times 10^{-19} \mbox{ ergs s}^{-1}\mbox{ cm}^{-2}\mbox{ \AA}^{-1})$ 
& 1.50(12) & 1.12(7) & 0.8(2) & 1.24(9)\\
$B_{0}$ (mag)\tablenotemark{b}  &26.58(8) & 26.88(7)  &
27.4(2) & 26.79(8)\\
\tableline
$\alpha_{\nu,2}$\tablenotemark{a} &\nodata  &\nodata & 2& 2\\
$F_{0,2}$ $( \times 10^{-19} \mbox{ ergs s}^{-1}\mbox{ cm}^{-2}\mbox{
  \AA}^{-1})$ & \nodata  &\nodata   & 0.6(2) & 0.27\\ 
$B_{0,2}$ (mag)\tablenotemark{b} &\nodata  &\nodata&  
27.5(3) & 28.5\\
\tableline
$\chi^{2}$\tablenotemark{c} & 1.9 & 26.5 & 0.8 & 1.4\\
DOF & 3 & 4 & 2& 3\\
$\chi^{2}/{\rm DOF}$ & 0.6 & 6.6 & 0.4 & 0.5 \\
\enddata
\tablenotetext{a}{The spectral index such that $F_{\lambda} \propto \lambda^{-(2+\alpha_{\nu})}$.}
\tablenotetext{b}{$B$-band Vega-magnitude corresponding to $F_{0}$.}
\tablenotetext{c}{The $\chi^{2}$ values are raw values that do not
  take into account the prior distributions (e.g., Eqn.~\ref{eqn:prior}).  They are there only as a
  guide, showing which models do and do not fit the data independent of the
  prior distributions.}
\tablecomments{Numbers in parentheses are 68\% confidence limits in
  the last digit(s).
  Values without confidence limits were held fixed for the fit.
  Values with subscript $0$ are for $\lambda_{0}=4500$~\AA.  $A_{0}$
  was constrained by use of Equation~\ref{eqn:prior}.}
\end{deluxetable}

In Tables~\ref{tab:fitnoA} and \ref{tab:fit} we give $\chi^{2}$ values
for each fit.  These values were computed from the data without taking
into account the prior distributions used in Table~\ref{tab:fit}.  The
best-fit values of the parameters were computed not by using the
$\chi^{2}$ values themselves buy through the marginalized likelihood
functions that incorporated the prior.  The $\chi^{2}$ values are
there only as a reference, to show that even with the prior
distribution in place the fits are still good. One should not use the
typical $\Delta \chi^{2}$ technique \citep[][p.\ 697]{numrec} for
determining parameter uncertainties --- while the PL models from
Tables~\ref{tab:fitnoA} and \ref{tab:fit} have $\chi^{2}$'s that
differ by 0.9, the PL model from Table~\ref{tab:fitnoA} is excluded
very significantly by the fit in Table~\ref{tab:fit}, as shown by the
small uncertainties in the parameters of Table~\ref{tab:fit}.  Instead
the confidence limits given in Tables~\ref{tab:fitnoA} and
\ref{tab:fit} are single-parameter 68\% limits determined from the
marginalized likelihood functions.

\begin{figure}[t]
\plotone{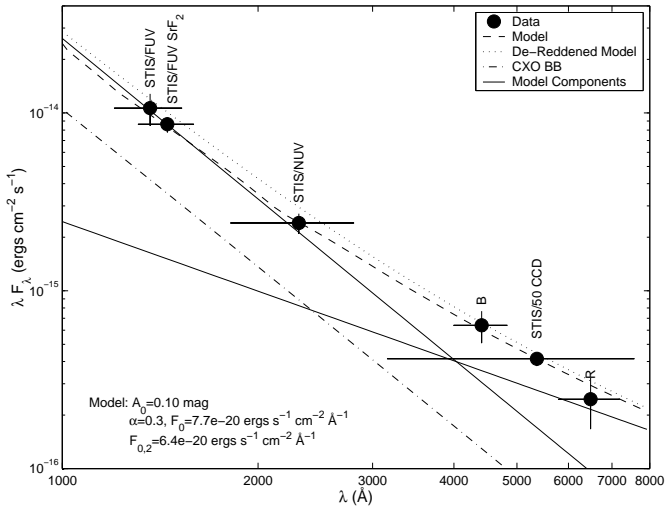}
\caption{Optical/UV spectrum of \rxj, with data from \citet{kvk98} and
this paper.  The PL+RJ fit (unconstrained Rayleigh-Jeans tail plus a
second PL) is shown: the best-fit model (dashed line), the best-fit
model corrected for extinction (dotted line), the two components of
the model (solid lines), and the extrapolation of the \chandra\
blackbody fit (dash-dotted line; ``CXO BB''; see \S~\ref{sec:xray}).
The different bands are labeled.}
\label{fig:spec2}
\end{figure}

We find a good fit for the single unconstrained power-law, shown in
Figure~\ref{fig:spec}.  This model has $\chi^{2}=1.9$ for 3
degrees-of-freedom (DOF).  The Rayleigh-Jeans fit, though, is
significantly worse, with $\chi^{2}=26.5$ for 4 degrees of freedom
under the same prior assumptions.  We can therefore reject
$\alpha_{\nu}=2$ with $>98$\% confidence.  For the PL+RJ fit, the
results are given in the PL+RJ column of Table~\ref{tab:fit}.  The
fit, shown in Figure~\ref{fig:spec2}, is good, with $\chi^{2}=0.8$ for
2 DOF.  For the final fit, PL+X-ray, we also find an acceptable value
of $\chi^{2}=1.4$ for 3 DOF.  We can conclude that all of the fits
except the RJ fit are acceptable.

As mentioned above, the values of $\langle \lambda \rangle$,
$A^{\prime}_{\lambda}$, the aperture correction and the zero-point
correction change depending on the spectral model.  However, except
for the RJ fit, all of the models have sufficiently similar flux
distributions, despite the different contributions from different
components, that the values are essentially the same for these models.
We give the values of these parameters for the best-fit values of
$\alpha_{\nu}$ and $A_{0}$ in Table~\ref{tab:obs}; these will apply to
the PL, PL+RJ, and PL+X-ray models.  The only filter whose calibration
changes significantly is the extremely broad-band STIS/50CCD.

As seen in Figure~\ref{fig:spec}, the PL fit has a shallower slope
than the extrapolation of the \chandra\ blackbody spectrum.  While an
extrapolation of the optical/UV PL does not intersect the X-ray
spectrum, it does come to within a factor of 1.2 (at $142\mbox{
\AA}$).  By the lower energy end of the \chandra\ data ($0.1\mbox{ keV}
\approx 125$~\AA), the optical/UV PL must have turned over as it is
not seen in the X-rays.  The power-law component of the PL+X-ray fit
behaves similarly.

\begin{figure}
\plotone{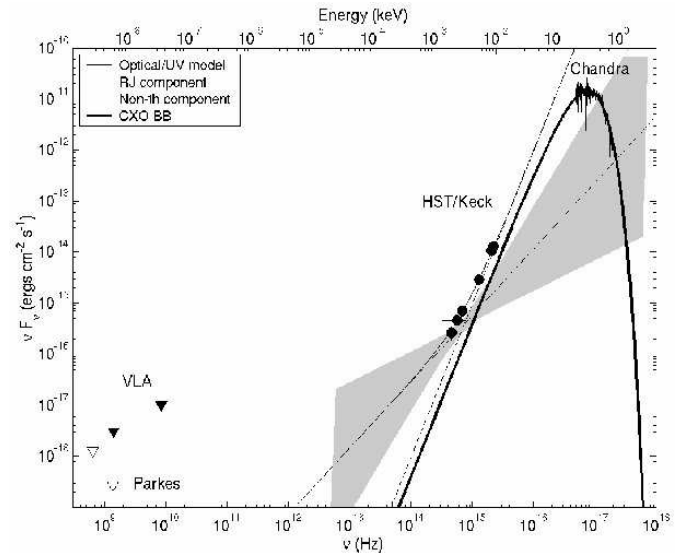}
\caption{ Broad-band spectrum of \rxj, from radio to X-rays.  The
absorption-corrected optical/UV data (\S~\ref{sec:opt}) are plotted as
filled circles, the absorption-corrected X-ray data
(\S~\ref{sec:xray}) as points, the VLA upper limits (\S~\ref{sec:vla})
as filled triangles, and the Parkes upper limits for pulse widths of
0.6--3\% (\S~\ref{sec:pulse}) as the open triangles.  The
models are: X-ray blackbody (thick solid line), Rayleigh-Jeans
component of the PL+RJ fit to the optical/UV data (dashed line), non-thermal
component of the PL+RJ fit (dash-dotted line), and
the overall PL+RJ fit to the optical/UV data (thin solid line).  The
$\pm1\,\sigma$ uncertainties on the non-thermal PL are shown by the shaded region.}
\label{fig:totalspec}
\end{figure}

The best-fit Rayleigh-Jeans component of the PL+RJ model is a factor
of 2.4(4) above the extrapolation of the X-ray blackbody
(Fig.~\ref{fig:spec2} and Tab.~\ref{tab:fit}).  The two components
contribute equally at $\lambda=4930\mbox{ \AA}$: at shorter
wavelengths the Rayleigh-Jeans component dominates, while at longer
wavelengths the non-thermal component dominates.  The non-thermal PL
component is above the X-ray extrapolation in the optical regime,
intersects it at 2680~\AA, and the continues below it.  Therefore, the
non-thermal PL would not be seen in soft X-rays or in the radio
(Fig.~\ref{fig:totalspec}), although it may approach the X-ray
spectrum at energies  $\gsim 2$~keV.

\subsubsection{Disk Fits}
We also considered fits to the optical/UV data that include a disk of
accreting material, such as that proposed to account for the X-ray
luminosity and periodicity of \rxj\ \citep[e.g.,][]{w97,kp97,alpar01}.
For the disk spectrum we used the model of \citet*{phn00}.  We found
fits using a disk model to be unsatisfactory.  As there are too many
free parameters to do a formal fit (the inner and outer disk radii,
the disk inclination, as well as two undetermined efficiency factors),
we varied subsets of the parameters by hand.  We first considered
disks that extend in to the corotation radius and out to infinity.
For most conceivable disk inclinations ($i \lsim 85\degr$) the disk
alone is a factor of 10--20 above the optical data, and this is
without any contribution to the optical emission from the neutron star
surface.  Toward the short-wavelength end of the disk, where the
emission decreases below the level of our data and the contribution is
primarily from the inner edge of the disk, the slope is entirely
inconsistent with the optical/UV data: it goes approximately as
$F_{\lambda} \propto \lambda^{4}$, while the excess flux in the
optical (compared to a Rayleigh-Jeans tail) is like $F_{\lambda}
\propto \lambda^{-2.3}$.  We therefore consider this disk model to be
very unlikely for \rxj.

There are disk models that can reproduce a spectrum roughly similar to that
observed.  This occurs when the inner radius is far inside the
corotation radius and approaches the neutron star surface, while the
outer radius move inward to $\sim 10^{8}$~cm.  But while the spectral shapes
are not inconsistent, the flux predicted by such disks is a factor of
$\sim 100$ above the optical/UV data.  Also, there is no natural
reason for the disk to be truncated at such small radii (the optical
data do not allow for any stellar companion).  Therefore
this disk model is also very unlikely for \rxj. 

\subsubsection{Variability}
It is possible that the spectrum of \rxj\ is a Rayleigh-Jeans tail in
the optical, and that the deviations we see are temporal in nature:
i.e.\ \rxj\ could vary.  However, we consider this unlikely.  First,
the X-ray flux has been extremely constant over almost a decade of
observation \citep{hmb+97,pmm+01}.  Second, similar sources such as
\rxjw\ and PSR~B0656+14 have exhibited constant optical fluxes, again
over several years of observations \citep{vkk01,kpz+01}.

Regardless, we can perform a simple test for variability.  We have an
ongoing series of \textit{HST} observations that, while designed to
measure the parallax and proper motion of \rxj, also provide a
sensitive flux monitor.  Only the first two epochs of data have been
observed so far (at MJDs 52459 and 52532).  The data are from 4950-s
observations with the Advanced Camera for Surveys on \textit{HST}
using the High Resolution Camera (ACS/HRC) in the F475W filter.  The
photometric calibration of ACS is not complete, so we cannot directly
compare the measured flux of \rxj\ to the models presented here
(although the new data are roughly consistent), but we can look for
variations in the flux of \rxj\ and in that of the other sources in
the field.  We drizzled the data using a preliminary model of the
ACS/HRC geometric distortion, and then performed aperture photometry
on \rxj\ and 11 other sources, ranging from much brighter than \rxj\
to about as bright as \rxj.  The field sources changed by $0.02\pm
0.04$~mag from the first epoch to the second, while \rxj\ changed by
$-0.10 \pm 0.09$~mag.  This number is preliminary, but shows that the
flux of \rxj\ changed by at most 10\% over two months.  We will
eventually have better-calibrated data spanning two years, which will
allow us to make a much more rigorous test for variability.

\section{X-ray Data}
\label{sec:xray}

\begin{deluxetable}{c c c c}
\tabletypesize{\small}
\tablecaption{Summary of \chandra\ HRC-S/LETG Observations\label{tab:cxo}}
\tablewidth{0pt}
\tablehead{
\colhead{Date} & \colhead{Exposure} & \mc{2}{c}{Counts} \\ \cline{3-4}
\colhead{(UT)} & \colhead{(ksec)} & \colhead{Order 0} & \colhead{Orders
  $\pm1$} \\
}
\startdata
2000-Feb-01 & 5.4 & 929 & 671 \\
2000-Feb-02 & 26.3 & 4584 & 3027 \\
2000-Feb-04 & 6.1 & 1119 & 687 \\
\enddata
\end{deluxetable}

\rxj\ was observed with the \textit{Chandra X-ray Observatory}, using the
High Resolution Camera spectroscopic detector (HRC-S) with the Low
Energy Transmission Grating (LETG); the observations are summarized in
Table~\ref{tab:cxo}.  Here we describe the spectroscopic analysis of
these data --- timing analysis is described in
\citet{kkvkm02} and \citet{zhc+02}.

\begin{figure}[b]
\plotone{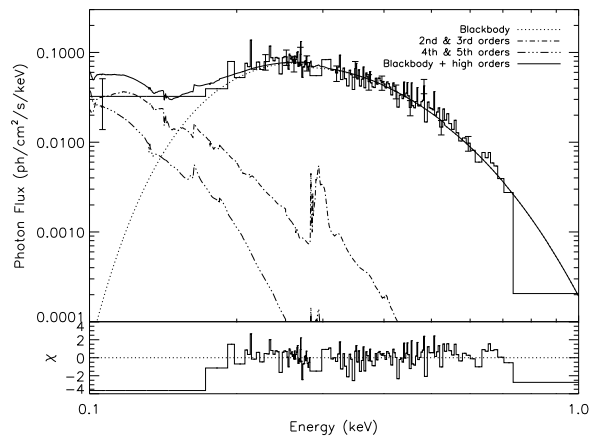}
\caption{LETGS spectrum of \rxj. The bin sizes have been varied to
provide good signals in each energy bin; the uncertainties are about
20\% everywhere. Solid line: Model consisting of one blackbody
component.  High orders do not
contribute significantly for $E > 0.20$~keV, while the data at low
energies ($E < 0.15$~keV) are best modeled as the result of the sum of
high orders.}
\label{fig:xspec}
\end{figure}

\begin{deluxetable}{c c c}
\tabletypesize{\small}
\tablecaption{One and Two Blackbody Fit to the LETG Spectrum\label{tab:xfit}}
\tablewidth{0pt}
\tablehead{
\colhead{Parameter} & \mc{2}{c}{Value}\\
}
\startdata
$N_{H}\;(\times 10^{20}\mbox{ cm}^{-2})$ & $1.32(14)$ & $1.46(14)$ \\
\tableline
$T_{\rm hot}\;(\times 10^{5}\mbox{ K})$ & $9.45(15)$ & $9.47(15)$ \\
$R_{\rm hot}\;(\mbox{km})$ & $6.1(3)d_{300}$ & $6.1(6)d_{300}$ \\
$T_{\rm cold}\;(\times 10^{5}\mbox{ K})$ & \nodata & $3.72(10)$ \\
$R_{\rm cold}\;(\mbox{km})$ & \nodata  & $15(17)$\\
\tableline
$\chi^{2}$ & 163.9 & 163.1 \\
DOF & 148 & 146\\
$\chi^{2}/{\rm DOF}$ & 1.11 & 1.12\\
\enddata
\tablecomments{Numbers in parentheses are 68\% confidence limits in
  the last digit(s).}
\end{deluxetable}

The spectral data were reduced from standard event lists using
\texttt{IDL} using custom processing scripts as described in
\citet{ms02} and \citet{mev+02}.  Raw events were extracted from
first and higher orders and calibrated with an updated model of the
LETGS effective area\footnote{This effective area is available at
  \url{http://cxc.harvard.edu/cal/Links/
Letg/User/Hrc\_QE/EA/correct\_ea/letgs\_NOGAP\_EA\_001031.mod}.} (EA), which was developed from observations
of PKS~2155$-$304..

Integrating the observed fluxes over the 0.25--3.0~keV band gives an
observed flux of $\expnt{(9\pm2)}{-12}\mbox{ ergs cm}^{-2}\mbox{
s}^{-1}$ and an absorbed luminosity of
$\expnt{(9\pm2)}{31}d_{300}^{2}\mbox{ ergs s}^{-1}$, where $d=300
d_{300}$~pc is the distance to \rxj\ \citep{kvka02}.  The data were
rebinned adaptively to provide a signal-to-noise ratio (S/N) of 5 in
each bin over the 0.10--2.0~keV range. The spectrum, shown in
Figure~\ref{fig:xspec}, was first estimated using the first-order EA
only. The contributions to the observed counts from high orders are
estimated by folding a model for first-order through the high-order EA
(important only below 0.2~keV).

Following previous analyses \citep{hmb+97,pmm+01}, we modeled the
continuum by an absorbed blackbody; the fitted parameters are given in
Table~\ref{tab:xfit}.  We exclude the data below 0.15~keV from the fit
where uncertainties in the high-order grating efficiencies can be
important.  This fit gives a temperature of 81.4(13)~eV and a
bolometric luminosity of $\expnt{2.1}{32}d_{300}^{2}\mbox{ ergs
s}^{-1}$, consistent with the \rosat\ and \xmm\ analyses
\citep{hmb+97,pmm+01},  given the uncertainties in modeling
the effective areas below 0.2~keV and the higher order responses.  The
fit has a reduced $\chi^{2}=1.11$ (for 148 DOF), acceptable at the 90\% level.
Other models, such as those with non-thermal power-laws or a second
blackbody, did not improve the fit.

\begin{figure}[t]
\plotone{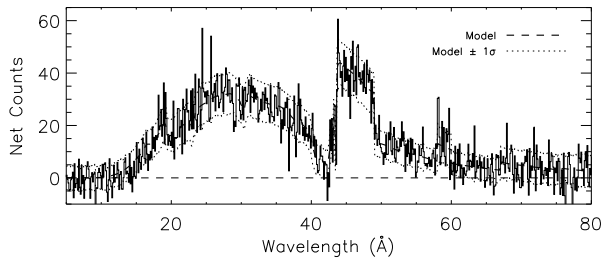}
\caption{Count spectrum of \rxj\ obtained with the LETGS. A binning of
0.125~\AA\ was used to obtain sufficient signals per bin to search for
narrow features. Heavy dashed line: Expected count spectrum from the
single blackbody 
model shown in Fig.~\ref{fig:xspec}. Light dotted lines: $\pm1\,\sigma$
uncertainties about the model. The residuals are consistent with
statistical fluctuations about the model. The sharp edges in the model
near the 50--70~\AA\ range are the result of detector gaps.}
\label{fig:xcountspec}
\end{figure}

\begin{figure}[b]
\plotone{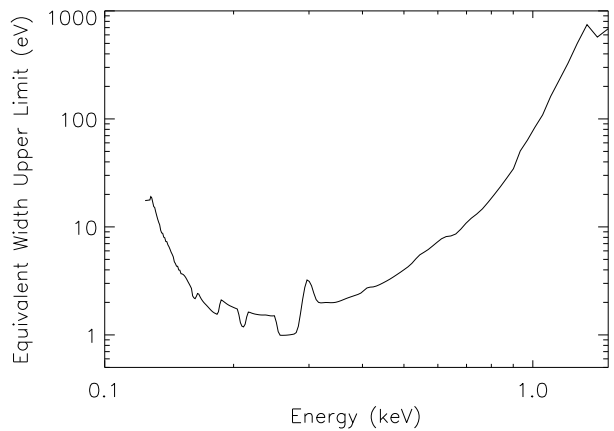}
\caption{Estimates of the $3\,\sigma$ limits that can be placed on any
emission lines whose FWHMs are comparable to the instrument
resolution, as a function of energy. The computation uses the model
fitted to the data (Fig.~\ref{fig:xspec}) and the effective areas. We
assume that candidate features are only 2 bins ($0.25$~\AA) wide. The
curves are rather smooth, except for locations of chip gaps, so one
may derive limits on broad features using these curves until the scale
of the feature becomes comparable to that of the instrument
calibration uncertainties. Limits on absorption features are identical
when there are many counts but are systematically larger at the high
and low ends of the spectrum, where there are fewer than 25 counts per
bin.}
\label{fig:ew}
\end{figure}

The best-fit model is plotted against the fitted data in
Figure~\ref{fig:xspec}. The count spectrum (Fig.~\ref{fig:xcountspec})
was binned at 0.125~\AA\ resolution in order to search for narrow
spectral features against the continuum model. No significant emission
or absorption features were found: in Figure~\ref{fig:ew} we give
$3\,\sigma$ upper limits to the equivalent width of narrow-line features in our
data.  The \chandra\ data are generally consistent with the \xmm\ data
\citep{pmm+01}, although now the upper limit for the energy  of any
narrow features is somewhat lower, around 0.2~keV.

\section{Radio Observations}
\label{sec:radio}
\subsection{Synthesis Imaging}
\label{sec:vla}
We observed \rxj\ with the Very Large Array (VLA) once at 8.4~GHz and
twice at 1.4~GHz (summarized in Table~\ref{tab:vla}) in the standard
synthesis-imaging mode.  All data sets
were independently calibrated using {\tt AIPS}, but the two 1.4~GHz
observations were combined for imaging.

\begin{deluxetable}{c c c c c c}
\tabletypesize{\small}
\tablecaption{Summary of VLA Observations\label{tab:vla}}
\tablewidth{0pt}
\tablehead{
\colhead{Date} & \colhead{Frequency} & \colhead{Exposure} & \colhead{Config.} &
\colhead{Beam Size} & \colhead{RMS}\\
\colhead{(UT)} & \colhead{(GHz)} & \colhead{(sec)} & & \colhead{(asec)} & \colhead{($\mu$Jy)}\\
}
\startdata
1998-Feb-07 & 8.4 & 6720 & D$\rightarrow$A\tablenotemark{a} & $15\arcsec \times 5\farcs8$ & 40\\
1999-Feb-18 & 1.4 & 4410 & DnC & $36\arcsec \times 32\arcsec$ & 80\\
1999-Apr-19\tablenotemark{b} & 1.4 & 7380 & D & \nodata & \nodata\\
\enddata
\tablecomments{Observations all had $2 \times 50$-MHz bandwidths.}
\tablenotetext{a}{Data were taken while switching from D configuration
  to A configuration.}
\tablenotetext{b}{Processed with the 1999-Feb-18 observation.}
\end{deluxetable}

For the 1.4~GHz data, we performed imaging and self-calibration in
{\tt difmap}.  We iteratively cleaned and self-calibrated (phase only)
until the gain solution converged. Uniform weighting was used,
yielding a synthesized beam with ${\rm FWHM}\approx 34\arcsec$.  An
overall gain adjustment was added for one IF, effectively correcting
for a non-zero spectral index across the two IF's.  No additional
amplitude self-calibration was necessary.

\begin{figure}[t]
\plotone{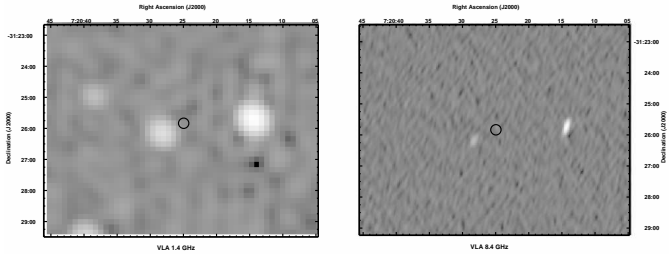}
\caption{VLA maps of the field around \rxj.  Left: 1.4~GHz map from
  the 1999 observations.  Right: 8.4~GHz map.  \rxj\  is
  indicated with the circle at the center (the radius of the circle is
  $10\arcsec$).}
\label{fig:vlazoom}
\end{figure}

After cleaning, we found rms map noise to be 0.08~mJy, a factor of
$\sim 8$ higher than the theoretical thermal noise but consistent
with confusion \citep{ccg+98}.  The final image (see
Fig.~\ref{fig:vlazoom}) shows a 5.0~mJy point source next to the
position of \rxj, but we believe that this source is unrelated.  For
reference, the radio source is at J2000 $\alpha=07^{\rm h}20^{\rm
m}28\fs28(2)$, $\delta= -31\degr26\arcmin09\farcs9(3)$, $46\arcsec$
away from the nominal position of the source.  No point-like or
diffuse emission from \rxj\ was found, which then gives $3\,\sigma$
upper limits to the flux of a point-source of 0.24~mJy and to that of
an extended source of $<0.43\mbox { mJy arcmin}^{-2}$.

The imaging of the 8.4~GHz data proceeded similarly.  Cleaning and
phase self-calibration were done in \texttt{difmap}.  Again, no source
was found at the position of \rxj\ (see Fig.~\ref{fig:vlazoom}), giving
a $3\,\sigma$ flux limit of $0.12$~mJy.

\subsection{Pulsation Searches}
\label{sec:pulse}
%

We observed RX~J0720$-$3125 with the 64-m Parkes radio telescope on 11
January 2001 in an attempt to detect pulsed radio emission.  The
target position was observed for 10,800~s at center frequencies of
1374~MHz and 644~MHz.  At the higher frequency we used the center beam
of the Parkes multibeam receiver to feed a $3\mbox{ MHz}\times
96$-channel multibeam filterbank (see \citealt{lcm+00}).  At 644~MHz
the front-end was the Parkes 50-cm receiver, and the back-end was a
$0.125\mbox{ MHz} \times 256$-channel filterbank.  Both observations
employed a 1-ms sample period and one-bit digitization.

The interstellar dispersion toward \rxj\ is unknown, but we can
estimate it with the latest model of Galactic electron density
\citep{cl02} which predicts a dispersion measure ${\rm DM}=4\mbox{ pc
cm}^{-3}$ at $d=300$~pc or ${\rm DM}=30\mbox{ pc cm}^{-3}$ at
$d=500$~pc.  We therefore take ${\rm DM}=100\mbox{ pc cm}^{-3}$ as a
conservative upper limit to the DM (the search was highly insensitive
to DM anyway, given the long period of \rxj).  Both datasets were
de-dispersed with dispersion measures up to $100\mbox{ pc cm}^{-3}$
and searched for periodicities near the known X-ray period
\citep{kkvkm02,zhc+02} using standard folding and FFT-based
techniques.  No pulsar-like signals were detected.

Without a detection, we must estimate the limiting flux of a signal.
While the search is insensitive to DM, the shape of the hypothetical
pulse profile strongly affects the sensitivity.  Assuming a pulse duty
cycle of $w=1$\%, our $8\,\sigma$ detection limits are 0.2~mJy at 644~MHz
and 0.02~mJy at 1374~MHz, and they  scale approximately as

\begin{eqnarray}
S_{\rm min,\,644\,MHz} & = & 2.0 \sqrt{\frac{w}{1-w}}\mbox{ mJy}\nonumber \\
S_{\rm min,\,1374\,MHz} & = & 0.18 \sqrt{\frac{w}{1-w}}\mbox{ mJy}.
\end{eqnarray}

For signals with high $w$ (i.e.\ few harmonics) the 1.4~GHz sensitivity is
comparable to that of the VLA observations (\S\ \ref{sec:vla}), where
the large bandwidth of the multibeam system compensates for its
smaller area and shorter integration.  However, for very narrow
signals (where many harmonics are summed) the periodicity search is a
factor of $\sim 7$ deeper than the VLA observations.

\section{Discussion}
\label{sec:discuss}

\subsection{The Spectrum}
We have shown in \S~\ref{sec:opt} that the optical/UV spectrum of
\rxj\ does not follow a pure Rayleigh-Jeans tail.  It is possible that
the emission (minus the contribution of the Rayleigh-Jeans tail of the
X-ray spectrum) is entirely non-thermal in origin (like the Crab pulsar,
where the non-thermal emission entirely overwhelms any thermal
component).  However, the slope of the spectrum, $\alpha_{\nu}=1.1$ is
much steeper than that seen for other pulsars, whole spectral indices
range from $\alpha_{\nu}=0.11$ for the Crab \citep{sll+00} to
$\alpha_{\nu}\sim -1$ for other sources \citep[][and references
therein]{zsk+02}.  In addition, the low spin-down power of \rxj\
($\dot E < \expnt{2.4}{31}\mbox{ ergs s}^{-1}$; \citealt{kkvkm02})
compared to sources like the Crab or even middle-aged radio pulsars
means that there is no reason to expect this much non-thermal emission
from \rxj.

Of the six other isolated neutron stars with good optical/UV data,
there are three (all $\lsim 10^{6}$~yr old and within 500~pc) that
show evidence for thermal optical emission: the pulsars PSR~B0656+14
\citep*{pwc97} and Geminga \citep*{mhs98}\footnote{\citet{mhs98} find
the PL+RJ model acceptable for Geminga, although a single non-thermal
PL is also allowed.  In addition, they require an absorption feature
in the spectrum.}, and the nearby isolated neutron star \rxjw\
\citep{vkk01}. [For the $10^{4}$-yr Vela pulsar, the optical emission
\citep{mc01} is dominated by the non-thermal component, but the X-ray
spectrum contains thermal and non-thermal contributions
\citep{pzs+01}.]  Of these sources, \rxjw\ and PSR~B0656+14 have
thermal X-ray spectra too, like \rxj.  We therefore believe that the
optical/UV emission from \rxj\ is mostly thermal in nature (i.e.\
$\alpha_{\nu}=2$), with the deviation from a Rayleigh-Jeans tail
arising either due to a multiplicative opacity or to an added
component.  We address each of these models separately.

\subsubsection{One Component Model with Absorption}
It is possible that the emission from \rxj\ is entirely thermal, but
that the underlying Rayleigh-Jeans tail is modified by a
frequency-dependent absorption to give the observed spectrum.  An
opacity $\kappa_{\nu} \propto \nu^{0.9}$ would give the correct
result.  However, this model is artificial, and is entirely contrary
to what is seen with \rxjw, PSR~B0656+14, and the X-ray spectrum of
\rxj\ itself (\S~\ref{sec:pl}).  While we cannot reject this model
based on our data alone, comparison with other sources makes it
unlikely.

\subsubsection{Two Component Model}
\label{sec:two}

\begin{figure}[b]
\plotone{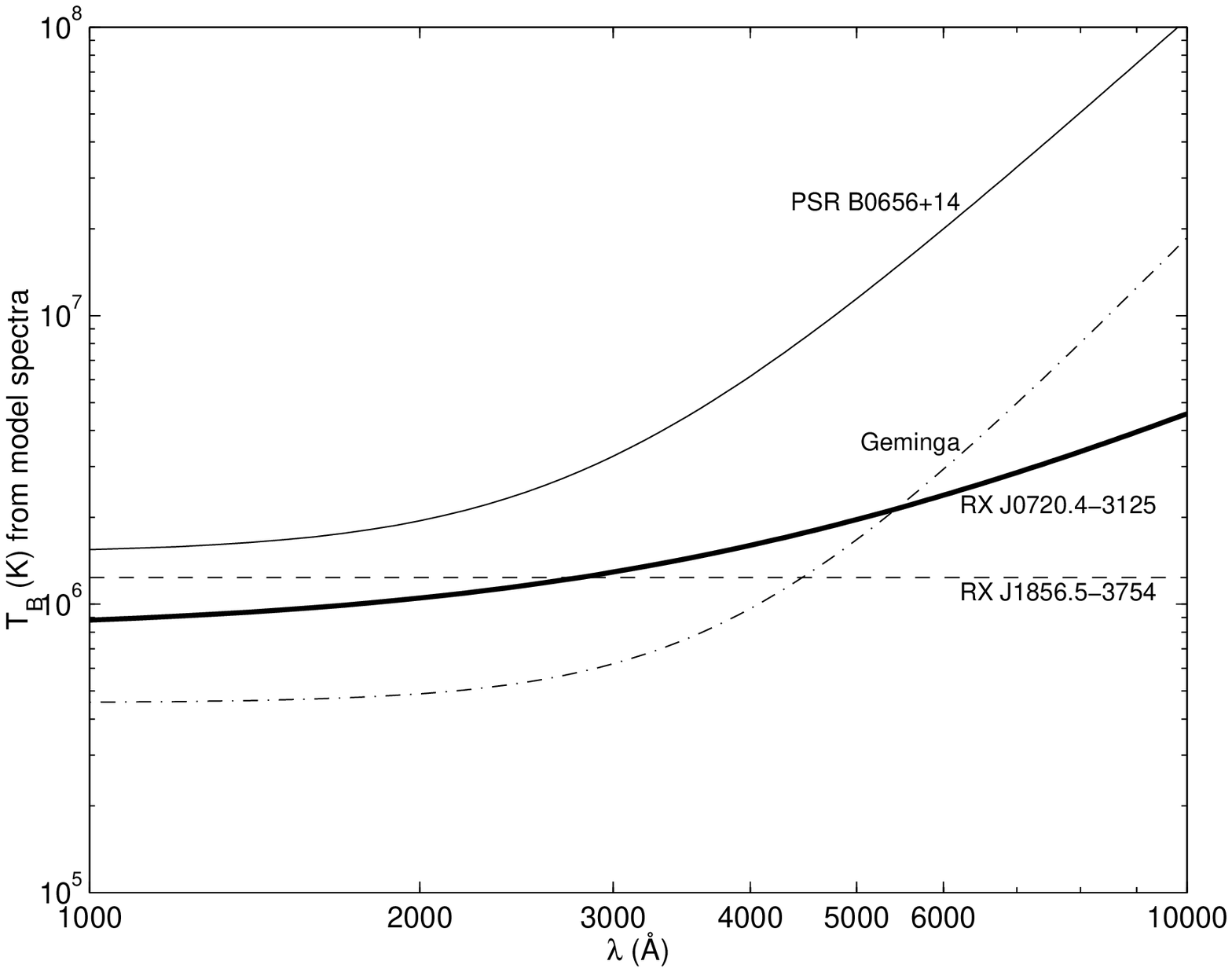}
\caption{Brightness temperature $T_{B}$  vs.\ wavelength for for \rxj\
  (solid line, this paper), \rxjw\ \citep[dashed line,][]{vkk01},
  PSR~B0656+14 \citep[dotted line,][]{pwc97}, and Geminga
  \citep[dot-dashed line,][]{mhs98}.  The brightness temperature is
  defined as $T_{B}(\lambda) \equiv\
  F_{\nu}(\lambda)\lambda^{2}\Omega/2k$, and the values were computed from
  model optical/UV spectra assuming a radius of 10~km and distances of
  300~pc, 140~pc, 330~pc \citep{kvka02,wl02}, and 160~pc
  \citep{cbmt96}, respectively.  Deviations from a constant $T_{B}$
  indicate non-thermal emission.}
\label{fig:allspec}
\end{figure}

A simpler and more physically motivated model for the optical/UV
spectrum of \rxj\ is that the emission is composed of significant
thermal emission plus a non-thermal PL (the PL+RJ model) similar to
the spectra of PSR~B0656+14 and Geminga.  We have plotted brightness
temperatures derived from model optical/UV spectra for these sources
and \rxjw\ in Figure~\ref{fig:allspec}.  The sources exhibit some
variety in their spectra, ranging from purely thermal (\rxjw) to
largely non-thermal (e.g., PSR~B0656+14).  While \rxj\ shows less
non-thermal emission than PSR~B0656+14 or Geminga, it is otherwise
unremarkable.

The (extinction corrected) $B$-band luminosity of \rxj\ is
$L_{B}=\expnt{6.5}{27}d_{300}^{2}\mbox{ ergs s}^{-1}$. If we separate
the non-thermal component, we find $L_{B,{\rm
non-th}}=\expnt{3.2}{27}d_{300}^{2}\mbox{ ergs s}^{-1}$.  This
compares well with the optical luminosities of similarly-aged pulsars
given in \citet{zsk+02}, which have $27 \lsim \log L_{B} \lsim 28$.
As such, \rxj\ fits in with the pulsar population despite its low
rotational power ($\dot E$).  We note, though, that all of the sources
in \citet{zsk+02} with $\tau \gsim 10^{5}$~yr have very similar values
of $L_{B}$, despite values of $\dot E$ that vary by about 3 orders of
magnitude, and that the values of $L_{B}$ show no apparent correlation
with $\dot E$.  So it appears that $L_{B}$ is not simply related to
$\dot E$, contrary to what is seen in the X-ray regime \citep[][see
below]{bt97}.  It may be, as suggested by \citet{zsk+02}, that the
efficiency of producing optical emission increases with time for $\tau
\gsim 10^{4}$~yr so that $L_{B}$ remains relatively constant despite
variations in $\dot E$ (i.e.\ $L_{B} = \eta(t) \dot E$, where
$d\eta(t)/dt > 0$ and $d\dot E/dt < 0$), or it may just be that the
$L_{B}$ is not directly coupled to $\dot E$.  A final possibility is
that the true distribution of $L_{B}$ is closely tied to $\dot E$ but
that we only see the brightest sources in this distribution due to
observational bias: a source with $\log L_{B}=26.5$ at 500~pc would
have $B\approx 31$~mag, below modern detection limits.  Similarly, the
very well studied \rxjw\ at $\approx 140$~pc, the limit to non-thermal
emission is $\log L_{B} \lsim 26.5$.  Therefore, any other source with
$\log L_{B} < 26.5$ could not be detected, and the true distribution
of $L_{B}$ may be poorly represented.

The slope of the non-thermal PL is $\alpha_{\nu} \approx 0.3$.  This
is consistent (within errors) with the slope of the optical PL for the
Crab, and is not that far from the slope of the spectrum of
PSR~B0656+14 ($\alpha_{\nu}=-0.45 \pm 0.26$; \citealt{kpz+01}).  The
non-thermal spectra of neutron stars have been observed to follow
roughly the same slope for over $\sim 5$ orders of magnitude in energy
\citep[e.g.,][]{kpz+01}, suggesting a single underlying mechanism for
the non-thermal optical and X-ray power-laws.  Here, the non-thermal
PL would contribute $\sim 1$\% of the flux of the blackbody in the
LETG band and would therefore be difficult to observe: $F(1\,{\rm
keV})=\expnt{(5^{+25}_{-4})}{-13}\mbox{ ergs s}^{-1}\mbox{ cm}^{-2}$.
At higher energies ($\gsim 2$~keV) the non-thermal PL could contribute
substantial X-ray flux (although the PL is highly uncertain at these
energies; see Figure~\ref{fig:totalspec}), flux that was not detected
in the \chandra\ or \xmm\ data.  The \xmm\ EPIC-pn data roughly give a
flux of $\expnt{2}{-14}\mbox{ ergs s}^{-1}\mbox{ cm}^{-2}$ at 1.5~keV,
consistent with a blackbody and barely consistent with the
extrapolated optical/UV power-law.  So, for \rxj\ the non-thermal
spectrum is not likely to bridge the X-ray and optical regimes.

\citet{bt97} have found a rough relation between
the spin-down power and the non-thermal X-ray emission of pulsars:
$L_{X,{\rm non-th}} \sim 10^{-3} \dot E$ (with $L_{X,{\rm non-th}}$ in
the \rosat\ band of 0.1--2.4~keV).  Applying this to \rxj\  we would
expect $L_{X,{\rm non-th}} \lsim 10^{28}\mbox{ ergs s}^{-1}$, or
$F_{X,{\rm non-th}} \lsim 10^{-15}\mbox{ ergs s}^{-1}\mbox{ cm}^{-2}$.
This is a factor of $>10^{4}$ less than the observed thermal X-ray
flux in the same band, and a factor of $10^{2}$ less than the
extrapolated non-thermal emission.  However, there is considerable uncertainty
in both the relation of \citet{bt97} (see, e.g., \citealt{pccm02}) and
in the extrapolation of the non-thermal PL to the X-ray regime, so the
difference between $F_{X,{\rm non-th}}$ predicted from the spin-down
and that predicted from the optical/UV spectrum may not be significant.

The thermal component in the optical/UV band is, like that of \rxjw,
above a simple extrapolation of the X-ray spectrum (for \rxjw\ it
exceeds the X-rays by a factor of $\sim 16$; \citealt{vkk01}).  This
could just be a matter of temperature, though.  For the sources
PSR~B0656+14 and PSR~B1055$-$52, the X-ray spectrum is best-fit by a
combination of a power-law and two blackbody components
(\citealt{gcf+96}; \citealt*{pwc97}; \citealt{kpz+01,ms02};
\citealt*{pzs02}), where the smaller hot portion (presumably the polar
cap) contributes the majority of flux in the traditional X-ray band
but the larger cool portion (along with the non-thermal emission)
gives rise to the optical/UV flux \citep{pzs02}.  In both cases, the
blackbody components are all hot enough ($\gsim \expnt{8}{5}$~K) to
appear in the X-ray band, and are therefore well modeled.

For \rxjw, which is closer in temperature to \rxj, \citet{br02} again
appealed to a multitemperature surface, but here the second blackbody
is not observed but only inferred (also see \citealt{wl02}).  It must
be too cool for the X-ray band ($T_{\rm cold} \lsim \expnt{5}{5}$~K),
while still hot enough to appear as a power-law in the optical/UV
band.  While not very well constrained, this model gave reasonable
results for \rxjw, including a weak constraint on the radius, and is
therefore valuable.   We now apply this model to \rxj.

We see from Table~\ref{tab:xfit} that a second (unconstrained)
blackbody component does not appreciably change the X-ray fit.  However, we
can use the goodness-of-fit (given by $\chi^{2}$) to constrain the
combinations of $T_{\rm cold}$ and $R_{\rm cold}$ that are allowed.
The flux of a Rayleigh-Jeans tail goes as $F_{\nu} \propto R^{2}T$.
We know that the X-ray blackbody has $R_{\rm hot}=6.1d_{300}$~km and $T_{\rm
hot}=\expnt{9.45}{5}$~K, and since the thermal fit to the optical/UV
data is a factor of 2.4 above the X-ray extrapolation, we find 
\beq
T_{\rm cold} = \expnt{2.4}{5}\left( \frac{R_{\rm
cold,15}}{d_{300}}\right)^{-2}\mbox{ K}, 
\eeq 
where the cold radius
$R_{\rm cold}=15 R_{\rm cold,15}$~km has been taken to be the radius
of the neutron star.  Such a cool blackbody would
not have been seen in the X-ray data (as shown by the large
uncertainties on $R_{\rm cold}$ in Tab.~\ref{tab:xfit}), but we 
can constrain $T_{\rm cold}$ to be $\approx (3.5$--$5.0)\times
10^{5}$~K, or $R_{\rm cold} \approx  (11$--$13)d_{300}$~km (at roughly 90\%
confidence).  This is similar to the 
temperature of the cold component found for \rxjw\ \citep{br02} ---
 not surprising since in both cases the cold component was
forced to give a Rayleigh-Jeans tail in the optical while not giving
significant contribution in the soft X-rays --- and the size agrees
well with estimates for the radius of a neutron star \citep{lp00}.

\citet{czr+01} model the phase-dependent hardness ratio of \rxj\ in
\xmm\ data, and find that it is consistent with a polar-cap model for
a large range of cap sizes with angular radii of $10$--$50\degr$.
This agrees quite well with our findings ($R_{\rm cold}=15d_{300}$~km
corresponds to an angular radius of $\approx 25\degr$).  

Taken together, these observations show that a RJ+PL model for the
optical/UV spectrum of \rxj\ makes it entirely consistent with being
an off-beam radio pulsar, one that likely has a cooler blackbody
component in the extreme UV.

\subsubsection{Constraints on the Magnetic Field}
We find no significant absorption features in the pulse-averaged or
pulse-phased spectra over the 0.15--0.80 keV band
(\S~\ref{sec:xray}). Thus, following \citet{ms02} and \citet{pmm+01},
we can rule out electron and proton cyclotron resonance lines in this
range.  \citet{pmm+01} already rule out the range $0.03 < B_{12} <
0.2$ and $50 < B_{12} < 200$, where $B_{12} = B/(10^{12}\;{\rm G})$.
By extending the spectrum down to 0.15~keV, we extend the lower limits
of the excluded ranges of magnetic fields to $B_{12}=0.015$ and
$B_{12}=25$, though the lower limits could increase if the bulk of the
emission comes from the equatorial zone, where the magnetic field is
$\sim 50$\% of the polar value, or if the absorbing plasma is far off
the neutron surface.  We can also use the lack of features in the
spectrum to exclude hydrogen atmospheres for a range of magnetic field
strengths.  Again following \citet{pmm+01}, we can exclude the range
$B_{12} > 15$.  The excluded ranges of $B$ agree with the finding that
\rxj\ is not a magnetar \citep{kkvkm02}, but otherwise the $B$ is
consistent with either of the models discussed in \citet[][see also
\S~\ref{sec:lum}]{kkvkm02}, namely that have $B_{12}\approx 1$--10.

\subsection{Radio Luminosity}
\label{sec:lum}
At a distance of $300 d_{300}$~pc, we limit the 1.4~GHz radio
luminosity of \rxj\ to $< \expnt{3}{25}d_{300}^{2}\mbox{ ergs
s}^{-1}$ ($L_{\rm rad}\equiv4\pi d^{2}F$), or following the radio
pulsar convention $L^{\prime}_{\rm rad}< 0.02\mbox{ mJy kpc}^{2}$
($L^{\prime}_{\rm rad}\equiv Fd^{2}$).  This is significantly below
what is expected of radio pulsars with parameters ($P$ and $\dot P$)
similar to those of \rxj.  For instance, the two high-$B$ radio
pulsars discovered by \citet{ckl+00} have $L_{\rm rad}\sim \expnt{5}{26}\mbox{
ergs s}^{-1}$, while the 8-s $10^{12}$-G pulsar \opsr\ has $L_{\rm rad}
\sim 10^{30}\mbox{ ergs s}^{-1}$ \citep{ymj99}.  Similarly, the radio
luminosity model of \citet*{acc02} predicts luminosities of
$10^{27-28}\mbox{ ergs s}^{-1}$, depending on the value of $\dot P$.
So we can see that \rxj, if it has any radio emission, must be beamed
away from the Earth.

If \rxj\ is like \opsr, then we might expect a similarly narrow radio
beam of $w \approx 0.6$\%, a beam width that agrees well with the
extrapolation of \citet{rankin93}.  \rxj\ may however be more similar
to the $10^{13}$~G radio pulsar \psr\ \citep{ckl+00}, which has a
significantly wider beam ($w\approx 3$\%).  In either case, we can
expect that the radio beam subtends a small solid angle, making the
lack of radio emission quite credible.  For such beams, our upper
limits to the radio luminosity \textit{decrease}, as the limit from
the Parkes data for a source with pulse width of $w=0.6$--$3$\% at
1.4~GHz is $L_{\rm rad}^{\prime}\approx 0.002\mbox{ mJy kpc}^{2}$.
Even the imaging (VLA) limit is quite faint, about a factor of 3
fainter than the limit for Geminga \citep{seiradakis92}, and a factor
of $\sim 30$ below that of PSR~J0205+6449 in 3C~58 \citep{csl+02}, but
the implied limit for a narrow pulse width is far below that of all
radio pulsars younger than $10^{6}$~yr \citep{motch01}.

With such small beams, the radio-quiet population of sources like
\rxj\ could potentially be very large, up to a hundred times the
radio-loud population (assuming a sharp cutoff in the radio beam).
While likely invisible to radio observations, such sources are of
course bright X-ray emitters, and would be visible to $\sim 5$~kpc in
a 30~ksec \xmm\ observation.  The total numbers of such sources
(either radio-loud or -quiet) are small, making statistics uncertain, but
there could be cooling radio-quiet neutron stars in as much as 1\% of
\xmm\ observations.  However, these sources would be all but
impossible to confirm, as there would be few X-ray photons for
spectral fitting or pulsation searches, and the optical/UV counterpart
would be extremely faint.  

\section{Conclusions}
\label{sec:conc}
We have shown, through a joint analysis of radio, optical, and X-ray
data, that the spectrum of the isolated neutron star \rxj\ cannot be
fit by a single blackbody model.  While statistically we cannot rule
out a model with a single power-law in the optical/UV domain and an
X-ray blackbody, from  a more general perspective we believe that
the best-fit model is one with three components: a hot ($\sim
\expnt{9}{5}$~K) blackbody on the polar cap, a cool ($\sim
\expnt{4}{5}$~K) blackbody over the whole surface, and a weak
non-thermal power-law in the optical/UV.  This is very similar to the
spectra of middle aged radio pulsars such as PSR~B0656+14 and
PSR~B1055$-$52, an observation that supports the identification of
\rxj\ with an off-beam radio pulsar.

\rxj\ appears extremely similar to the very nearby \rxjw, perhaps with
orientation being the only difference between them \citep{br02}.  We
believe it likely that non-thermal emission and/or pulsations will be
detected eventually for \rxjw, as suggested by \citet{br02}.  Both
sources seem to have spectra primarily composed of featureless
blackbodies.  If we can develop a full understanding of such spectra
to properly relate the blackbody radii (such as that given here) to
the true radii, these sources will be ideal targets for the
determination of the equation-of-state.

In the near future, we will obtain \hst\ astrometry allowing us to
determine the distance and velocity of \rxj, as well as H$\alpha$
imaging to search for bow-shock nebulae \citep[e.g.,][]{vkk01b} that
will place significant constraints on alternate models such as
accretion.  In the off-beam pulsar model, these data may limit the
luminosity of any particle wind from \rxj, and will reduce uncertainty
in the EOS by determining the conversion from solid angle to radius.

\acknowledgements We thank the anonymous referee for valuable
comments.  D.~L.~K.\ is supported by the Fannie and John Hertz
Foundation and S.~R.~K.\ by NSF and NASA.  M.~H.~v.~K.\ is supported
by a fellowship from the Royal Netherlands Academy of Arts and
Sciences.  H.~L.~M.\ was supported under NASA contract SAO SV1-61010.
DLK also thanks \chandra\ grant GO0-1024X for additional support.
Data presented herein were based on observations made with the
NASA/ESA Hubble Space Telescope, obtained at the Space Telescope
Science Institute, which is operated by the Association of
Universities for Research in Astronomy, Inc., under NASA contract NAS
5-26555.  The National Radio Astronomy Observatory is a facility of
the National Science Foundation operated under cooperative agreement
by Associated Universities, Inc.  Data presented herein were also
obtained at the W.~M.~Keck Observatory, which is operated as a
scientific partnership among the California Institute of Technology,
the University of California, and the National Aeronautics and Space
Administration.  The Guide Star Catalog-II is a joint project of the
Space Telescope Science Institute and the Osservatorio Astronomico di
Torino.

\bibliographystyle{apj}


\end{document}